\documentclass[11pt]{article}
\usepackage{graphicx}
\usepackage{cite}
\usepackage{epstopdf}
\usepackage{amssymb}
\usepackage{epstopdf}
\usepackage{hep}
\usepackage{feynmp}
\usepackage{multicol}
\usepackage{simplewick}
\usepackage{appendix}
\usepackage{subfigure}
\newcommand{\eins}{\mbox{$1 \hspace{-1.0mm} {\bf l}$}}

\begin{document}

\title{{\bf Fragmentation contribution to the transverse single-spin asymmetry in proton-proton collisions}}

\author{A.~Metz and D.~Pitonyak
 \\[0.3cm]
{\normalsize\it Department of Physics, Barton Hall,
  Temple University, Philadelphia, PA 19122, USA} \\[0.15cm]
}

\date{\today}
\maketitle

\begin{abstract}
\noindent
Within the collinear twist-3 framework, we study the single-spin asymmetry (SSA) in collisions between unpolarized protons and transversely polarized protons with focus on the fragmentation term.  The fragmentation mechanism must be analyzed in detail in order to unambiguously determine the impact of various contributions to SSAs in hadron production.  Such a distinction may also settle the ``sign mismatch'' between the transverse SSA in proton-proton collisions and the Sivers effect in semi-inclusive deep inelastic scattering.  We calculate terms involving quark-quark and quark-gluon-quark correlators, which is an important step in such an investigation.
\end{abstract}

\allowdisplaybreaks

%
%
%
\section{Introduction}
\label{s:intro}
Beginning in the late 1970s, calculations of transverse SSAs in inclusive hadron production within the na\"{i}ve collinear parton model demonstrated that such asymmetries should be on the order of $\alpha_{s}m_{q}/P_{h\perp}$, where $m_{q}$ is the mass of the quark, and $P_{h\perp}$ is the transverse momentum of the detected hadron \cite{Kane:1978nd, Ma:2008gm}.  These predictions contradict the large SSAs seen in experiments \cite{Bunce:1976yb, Adams:1991rw, Krueger:1998hz, Adams:2003fx, Adler:2005in, :2008mi, Adamczyk:2012xd}.  However, the use of  collinear twist-3 multi-parton correlators established a framework, valid when $\Lambda_{QCD} \ll P_{h\perp}$, that could potentially handle these observables \cite{Ellis:1982wd,Efremov:1981sh, Qiu:1991pp, Qiu:1998ia}.  Various processes have been analyzed over the last two decades using this methodology --- see \cite{Qiu:1991pp, Qiu:1998ia, Eguchi:2006qz, Kouvaris:2006zy, Koike:2007rq, Vogelsang:2009pj, Zhou:2009jm, Kanazawa:2010au, Koike:2009ge, Gamberg:2012iq, Metz:2012ui, Schlegel:2012ve, Kang:2012ns, Kanazawa:2012kt} for some specific examples.  (We also mention that other mechanisms have been proposed to explain large SSAs \cite{Hoyer:2006hu, Qian:2011ya, Kovchegov:2012ga, Troshin:2012fr}.)  Furthermore, the collinear twist-3 formalism has been used to describe the double-spin observable $A_{LT}$ in different reactions with one large scale \cite{Jaffe:1991kp, Tangerman:1994bb, Koike:2008du, Lu:2011th, Metz:2010xs, Kang:2011jw, Liang:2012rb, Metz:2012fq}.

In particular, much attention has been given to the transverse target SSA for inclusive single hadron production in proton-proton collisions.  This observable was initially believed to be dominated by soft-gluon-pole (SGP) contributions on the side of the transversely polarized proton \cite{Qiu:1998ia,Kouvaris:2006zy}, which involves the Efremov-Teryaev-Qiu-Sterman (ETQS) function $T_{F}(x,x)$.  However, a recent fit of $T_F(x,x)$ to $p^\uparrow p\rightarrow \pi X$ SSA data based on this assumption \cite{Kouvaris:2006zy} has led to the so-called ``sign mismatch'' crisis~\cite{Kang:2011hk} involving the ETQS function and the transverse momentum dependent (TMD) Sivers function extracted from semi-inclusive deep inelastic scattering (SIDIS) \cite{Anselmino:2008sga} --- see also the discussion in \cite{Metz:2012ui}.   A conceivable resolution could be that SFPs and/or tri-gluon correlations in the proton provide a significant effect.  However, even when one calculates the former \cite{Koike:2009ge} and includes them in a fit of $T_{F}(x,x)$ \cite{Kanazawa:2010au}, that function has the same sign as the one extracted in Ref.$\!$\cite{Kouvaris:2006zy}.  Also, tri-gluon correlations seem to only be important for processes dominated by gluon-gluon or photon-gluon fusion (e.g., asymmetries in $J/\psi$ production), and for pion production will probably only be relevant in the small and negative $x_F$ regions \cite{Beppu:2010qn,Koike:2011mb}.  Therefore, neither SFPs nor tri-gluon correlations seem likely to settle the sign mismatch issue.  In addition, chiral-odd collinear twist-3 functions on the side of the unpolarized proton were shown to give insignificant contributions \cite{Kanazawa:2000hz}, and, therefore, cannot help us with this matter.  Other possible explanations, like nodes in $x$ or $k_\perp$ in the Sivers function, have also been explored \cite{Kang:2011hk, Kang:2012xf}, but also seem unable to resolve the crisis.  One important term that remains is the fragmentation mechanism, which could at the very least give a piece comparable to the SGP contributions --- see \cite{Kang:2010zzb, Kang:2011ni, Anselmino:2012rq} and references therein --- and may be able to account for the sign mismatch. 

Actually, the fragmentation contribution has a counterpart in the TMD factorization approach known as the Collins mechanism \cite{Collins:1992kk}.  This causes azimuthal modulations in the cross section for processes like SIDIS and (almost back-to-back) di-hadron production from electron-positron annihilation.  These effects have been measured in both processes \cite{Airapetian:2004tw, Abe:2005zx, Alexakhin:2005iw, Garzia:2012za, Qian:2011py}, which has allowed for an extraction of the Collins function \cite{Vogelsang:2005cs, Efremov:2006qm, Anselmino:2007fs}.  In addition, both the collinear twist-3 and TMD approaches to SSAs from fragmentation in SIDIS have been shown to agree in an intermediate momentum region where both formalisms are valid \cite{Yuan:2009dw}.  The TMD Collins mechanism has also been used to describe the fragmentation piece of the SSA in $p^\uparrow p\rightarrow \pi X$ \cite{Kang:2011ni, Anselmino:2012rq}, although a rigorous proof has not been put forth that such a framework can be applied to a process with one large scale.

Within the collinear twist-3 approach, attempts have been made to formulate the fragmentation term in the SSA for inclusive pion production from proton-proton collisions \cite{Koike:2002ti}.  However, further work determined the contribution considered in Ref.~\cite{Koike:2002ti} vanishes due to a universality argument \cite{Metz:2002iz, Collins:2004nx, Gamberg:2008yt, Yuan:2007nd, Meissner:2008yf}.  Recent calculations have rectified the situation, and the so-called derivative term was computed for the first time in Ref.~\cite{Kang:2010zzb}.  In the current work, we derive not only this term but also the non-derivative term as well as contributions involving quark-gluon-quark ($qgq$) correlators.  In addition, we briefly comment on the implications of this and future studies on the resolution of the ``sign mismatch'' puzzle as well as the overall understanding of SSAs in proton-proton collisions.

The Letter is organized as follows:~in Sec.~\ref{s:T3form} we review the collinear twist-3 formalism, and, in particular, discuss the relevant unpolarized fragmentation functions (FFs).  In Sec.~\ref{s:calccsFrag} we present the result for the fragmentation contribution to the single-spin dependent cross section in $p^\uparrow p\rightarrow h X$ and give a few details of the calculation.  We conclude the Letter in Sec.~\ref{s:sum}.

%
%
%
\section{Collinear twist-3 formalism and unpolarized FFs}
\label{s:T3form}
To start, let us make explicit the process under consideration, namely,
\begin{equation}
A(P,\,\vec{S}_{\perp}) + B(P') \rightarrow C(P_{h}) + X\,, \label{e:process}
\end{equation}
where the 4-momenta and polarizations of the incoming protons $A$, $B$ and outgoing hadron $C$ are indicated.  The Mandelstam variables for the process are defined as $S = (P+P')^{2}$, $T = (P-P_h)^{2}$, and $U = (P'-P_h)^{2}$, which on the partonic level give $\hat{s} = xx' S$, $\hat{t} = xT/z$, and $\hat{u} = x' U/z$.  The longitudinal momentum fraction $x$ ($x'$) is associated with partons in the transversely polarized (unpolarized) proton.  In analogy to the usually defined lightcone vectors $n=(0^{+},1^{-},0_{\perp})$ and $\bar{n}=(1^{+},0^{-},0_{\perp})$, we also have $n_{h} \sim P_{h}$ and $\bar{n}_{h} \sim (P_{h}^{0}, -\vec{P}_{h})$ (with $n_{h}\cdot\bar{n}_{h} = 1$) as lightcone vectors associated with the outgoing hadron's direction of motion \cite{Kang:2010zzb}.  Such vectors allow us to perform the appropriate twist expansion of the fragmentation correlator in the context of this process.  We also note that $\epsilon_{\perp}^{\mu\nu} = \epsilon^{\rho\sigma\mu\nu}\bar{n}_{\rho}n_{\sigma}$ with $\epsilon_\perp^{12} = 1$.  We perform the calculation of the transverse SSA in the proton-proton {\it cm}-frame, with the transversely polarized proton moving along the positive $z$-axis.

The first non-vanishing contribution to the spin-dependent cross section is given by terms of twist-3 accuracy and reads 
\begin{align} 
d\sigma(\vec{P}_{h\perp},\vec{S}_{\perp}) &= \,H\otimes f_{a/A(3)}\otimes f_{b/B(2)}\otimes D_{C/c(2)} \nonumber \\*
&+ \,H'\otimes f_{a/A(2)}\otimes f_{b/B(3)}\otimes D_{C/c(2)} \nonumber \\
&+ \,H''\otimes f_{a/A(2)}\otimes f_{b/B(2)}\otimes D_{C/c(3)}\,,
\label{e:collfac}
\end{align} 
where a sum over partonic channels and parton flavors in each channel is understood.  In Eq.~(\ref{e:collfac}), $f_{a/A(t)}$ ($f_{b/B(t)}$) denotes the distribution function associated with parton $a$ ($b$) in proton $A$ ($B$), while $D_{C/c(t)}$ represents the fragmentation function associated with hadron $C$ in parton $c$.  The twist of the functions is indicated by $t$.  The factors $H$, $H'$, and $H''$ give the hard parts corresponding to each term, while the symbol $\otimes$ denotes convolutions in the appropriate momentum fractions.  As mentioned, the first term in (\ref{e:collfac}) has been analyzed previously in the literature \cite{Qiu:1998ia, Kouvaris:2006zy, Koike:2007rq, Koike:2009ge}.  Likewise, the second term, which involves chiral-odd twist-3 unpolarized distributions, has also been studied and was shown to be negligible because of the smallness of the hard scattering coefficients \cite{Kanazawa:2000hz}.  The focus of this work will be on the third term in Eq.~(\ref{e:collfac}) involving collinear twist-3 FFs.  Therefore, for the situation we consider, $f_{a/A(2)} = h_{1}^{a}$ and $f_{b/B(2)} = f_{1}^{b}$, where $h_{1}$ and $f_{1}$ are the standard twist-2 transversity distribution function and unpolarized distribution function, respectively.  We then must determine what contributions are possible for $D_{C/c(3)}$.

\begin{figure}[t]
\begin{center}
\includegraphics[width=14cm]{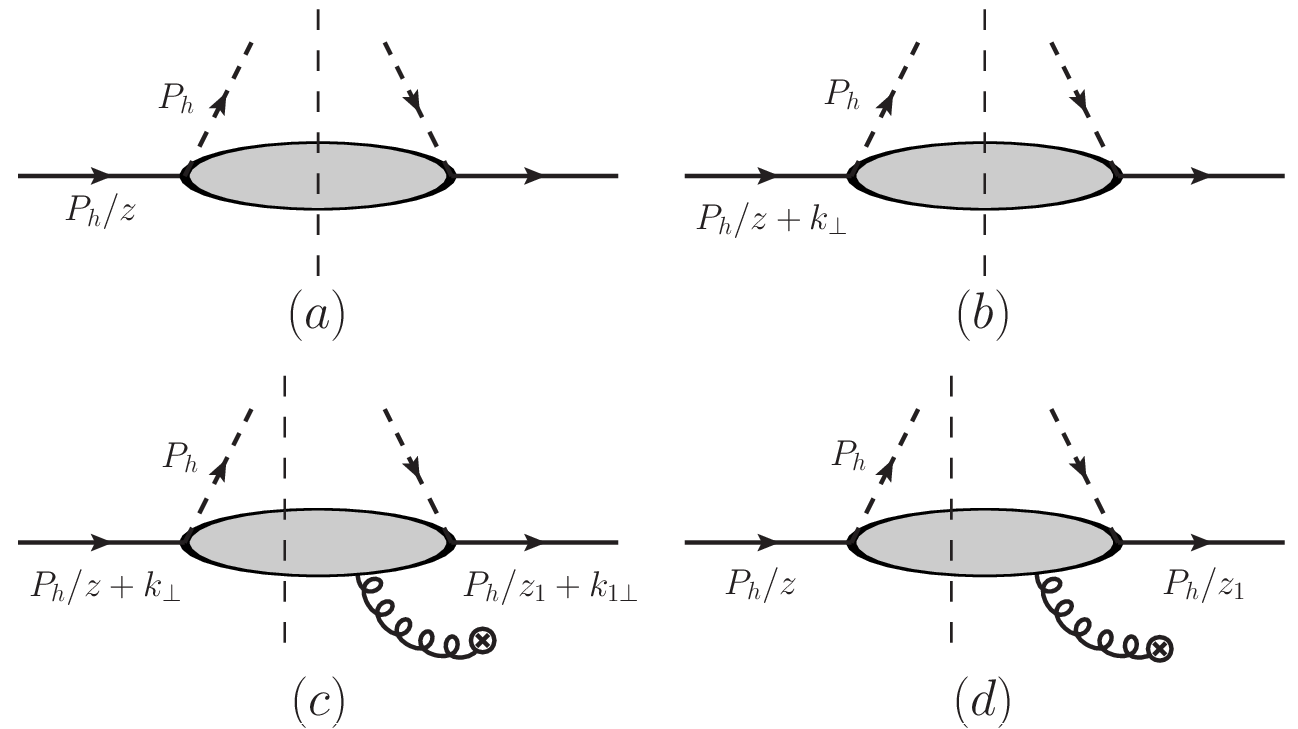}
\caption[]{Feynman diagrams for the twist-3 matrix elements that give contributions to $D_{C/c(3)}$.  See the text for more details.}
 \label{f:T3matrixFrag}
\end{center}
\end{figure}

For a detailed discussion of collinear twist-3 parton distribution functions (PDFs) see, e.g., \cite{Zhou:2009jm}; the same formalism can be generalized to the fragmentation case.  Here we consider the situation where the outgoing hadron has a large minus-component of momentum.  The twist-3 matrix elements that we must consider are given by the diagrams in Fig.~\ref{f:T3matrixFrag}.  (Note that tri-gluon correlators are only relevant for fragmentation into a transversely polarized hadron.)  In the lightcone ($A^- = 0$) gauge, these graphs lead to the three matrix elements
\begin{equation}
\langle\psi|\,\rangle\langle\,|\bar{\psi}\rangle\,, \;\langle\partial_{\perp}\psi|\,\rangle\langle\,|\bar{\psi}\rangle\,,\;\langle A_{\perp}\psi |\,\rangle\langle\,|\bar{\psi}\rangle\,,
\end{equation}
which result from Figs.~\ref{f:T3matrixFrag}(a), (b), and (d), respectively.  The symbol $|\,\rangle\langle\,|$ represents the intermediate $|P_{h};X\rangle\langle P_{h};X|$ in the fragmentation correlators.  We do not have to consider Fig.~\ref{f:T3matrixFrag}(c) because one does not need to simultaneously take into account $k_\perp$ expansion and $A_{\perp}$ gluon attachments (which would give rise to twist-4 contributions).

Now that we have determined the relevant twist-3 matrix elements, we must parameterize them in terms of twist-3 FFs that will eventually be involved in our final result.  We will follow the work of Ref.~\cite{Meissner:2009} in defining these FFs and the relations between them.  We first focus on the $qgq$ matrix element $\langle A_{\perp}\psi |\,\rangle\langle\,|\bar{\psi}\rangle$.  One notices that this matrix element is not gauge invariant.  In analogy to the $qgq$ distribution correlators, this can be resolved in two ways:~rewrite the gluon field $A_{\perp}$ in terms of the field strength tensor $F^{-\mu}_{\perp} = \partial^{-}A_{\perp}^{\mu}$ or rewrite it in terms of the covariant derivative $D_{\perp}^{\mu} = \partial_{\perp}^{\mu}-igA_{\perp}^{\mu}$.  Thus, we also have so-called ``F-type'' and ``D-type'' functions on the fragmentation side.  For unpolarized final-state hadrons, these read
\begin{eqnarray}
&&\sum_{X}\hspace{-0.55cm} \int \; \frac{1} {z}\int \frac{d\xi^{+}} {2\pi}\int \frac{d\zeta^{+}} {2\pi} e^{i\frac{P_{h}^{-}} {z_{1}}\xi^{+}} e^{i\left(\frac{1} {z}-\frac{1} {z_1}\right)P_{h}^{-}\zeta^{+}} \langle 0|igF_{\perp}^{-\mu}(\zeta^{+})\psi_{\alpha}(\xi^{+})|P_{h};X\rangle\langle P_{h};X|\bar{\psi}_{\beta}(0)|0\rangle \nonumber \\
&& \hspace{0.5cm} = M_{h}\left[\epsilon_{\perp}^{\mu\nu}\,\sigma_{\nu}^{\;\,+}\gamma_{5}\,\hat{H}^q_{FU}(z,z_{1})\right]_{\alpha\beta},
\label{e:F-typeFF}
\end{eqnarray}
and
\begin{eqnarray}
&&\sum_{X}\hspace{-0.55cm} \int \; \frac{1} {z}\int \frac{d\xi^{+}} {2\pi}\int \frac{d\zeta^{+}} {2\pi} e^{i\frac{P_{h}^{-}} {z_{1}}\xi^{+}} e^{i\left(\frac{1} {z}-\frac{1} {z_1}\right)P_{h}^{-}\zeta^{+}} \langle 0|iD_{\perp}^{\mu}(\zeta^{+})\psi_{\alpha}(\xi^{+})|P_{h};X\rangle\langle P_{h};X|\bar{\psi}_{\beta}(0)|0\rangle \nonumber \\
&& \hspace{0.5cm} = \frac{M_{h}} {P_{h}^{-}}\left[\epsilon_{\perp}^{\mu\nu}\,\sigma_{\nu}^{\;\,+}\gamma_{5}\,\hat{H}^q_{DU}(z,z_{1})\right]_{\alpha\beta}.
\label{e:D-typeFF}
\end{eqnarray}
We remark that these functions contain both real and imaginary parts, which we respectively indicate by $\hat{H}^{q,\Re}_{FU(DU)}(z,z_{1})$ and $\hat{H}^{q,\Im}_{FU(DU)}(z,z_{1})$. In Eqs.~(\ref{e:F-typeFF}), (\ref{e:D-typeFF}), we have suppressed Wilson lines and have indicated the hadron mass by $M_h$.  Moreover, the F-type and D-type functions are not independent of each other.  One can establish the following relation between them:
\begin{equation}
\hat{H}^q_{DU}(z,z_{1}) = - \frac{i} {z^2} \hat{H}^q(z)\,\delta\left(\frac{1} {z}-\frac{1} {z_{1}}\right) + PV \frac{1} {\frac{1} {z}-\frac{1} {z_{1}}}\hat{H}^q_{FU}(z,z_{1})\,,
\label{e:DHhatF}
\end{equation}
where $PV$ denotes the principal value.  In order to derive this expression, notice that we must introduce an additional twist-3 function $\hat{H}^q(z)$, which is given by
\begin{eqnarray}
&&\sum_{X}\hspace{-0.55cm}\int \; z\int\frac{d\xi^{+}} {2\pi} e^{i\frac{P_{h}^{-}} {z}\xi^{+}}\langle 0 |\left( iD_\perp^\mu(\xi^{+}) + g\int_{\xi^{+}}^\infty d\zeta^+ F_\perp^{-\mu}(\zeta^+) \right)\psi_{\alpha}(\xi^{+})|P_{h}; X\rangle\langle P_{h}; X|\bar{\psi}_{\beta}(0)|0\rangle \nonumber\\
&&\hspace{0.5cm} = -iM_{h}\left[\epsilon_{\perp}^{\mu\nu}\,\sigma_{\nu}^{\;\,+}\gamma_{5}\,\hat{H}^q(z)\right]_{\alpha\beta}.
\end{eqnarray}
This function is associated with the quark-quark ($qq$) matrix element $\langle\partial_{\perp}\psi|\,\rangle\langle\,|\bar{\psi}\rangle$.  We also mention that $\hat{H}(z)$ is equivalent to the first $k_{\perp}$-moment of the Collins function $H_{1}^{\perp}(z,z^2\vec{k}_{\perp}^{2})$ (as defined in \cite{Mulders:1995dh, Bacchetta:2006tn}) for the fragmentation of a transversely polarized quark into an unpolarized hadron:
\begin{equation}
\hat{H}^q(z) = z^2\int d^2 \vec{k}_{\perp} \, \frac{\vec{k}_{\perp}^{\,2}}{2 M_h^2} \, 
H_{1}^{\perp \hspace{0.025cm}q}(z,z^2\vec{k}_{\perp}^{\,2})\,.
 \label{e:H1perp}
\end{equation}
We note that $\hat{H}^q(z)$ is a real valued function, so that Eq.~(\ref{e:DHhatF}) in particular tells us
\begin{align}
\hat{H}^{q,\Im}_{DU}(z,z_{1}) &= - \frac{1} {z^2} \hat{H}^q(z)\,\delta\left(\frac{1} {z}-\frac{1} {z_{1}}\right) + PV \frac{1} {\frac{1} {z}-\frac{1} {z_{1}}}\hat{H}^{q,\Im}_{FU}(z,z_{1})\,,\label{e:DHhatFIm} \\*[0.3cm]
\hspace{1cm}\hat{H}^{q,\Re}_{DU}(z,z_{1}) &= PV \frac{1} {\frac{1} {z}-\frac{1} {z_{1}}}\hat{H}^{q,\Re}_{FU}(z,z_{1})\,. \label{e:DHhatFRe}
\end{align}
The other relevant $qq$ matrix element $\langle\psi|\,\rangle\langle\,|\bar{\psi}\rangle$ leads to contributions from the twist-3 functions $H^q(z)$, $E^q(z)$, whose definitions are given by \cite{Mulders:1995dh}
\begin{eqnarray}
\sum_{X}\hspace{-0.55cm}\int \; z \int\frac{d\xi^{+}} {2\pi} e^{i\frac{P_{h}^{-}} {z}\xi^{+}}\langle 0 |\psi_{\alpha}(\xi^{+})|P_{h}; X\rangle\langle P_{h}; X|\bar{\psi}_{\beta}(0)|0\rangle = \frac{M_{h}} {2P_{h}^{-}}\left[-i\epsilon_{\perp}^{\mu\nu}\,\sigma_{\mu\nu}\gamma_{5}\,H^q(z) + 2E^q(z)\cdot \eins\right]_{\alpha\beta}.
\end{eqnarray}
However, $H^q(z)$ ($E^q(z)$) can be related to the imaginary (real) part of the D-type function through the QCD equations of motion (EOM):
\begin{align} 
H^q(z) &= 2z^3\int \frac{d z_1} {z_1^2} \hat{H}^{q,\Im}_{DU}(z,z_{1})\,, \label{e:EOMH} \\[0.3cm]
E^q(z) &= -2z^3\int\! \frac{d z_1} {z_1^2} \hat{H}^{q,\Re}_{DU}(z,z_{1})\,. \label{e:EOME}
\end{align}
Thus, we see from Eqs.~(\ref{e:DHhatF}), (\ref{e:EOMH}), (\ref{e:EOME}) that of the five collinear twist-3 FFs relevant for an unpolarized hadron ($\hat{H}, H, E, \hat{H}_{FU}, \hat{H}_{DU}$), only two are independent of each other:~we can either work with $\hat{H},\hat{H}_{FU}$ or $\hat{H},\hat{H}_{DU}$.  However, we will find it convenient to write the final result in terms of $\hat{H}$, $H$, and $\hat{H}_{FU}$.  In fact, it turns out only $\hat{H}^{\Im}_{FU}$ is relevant for the calculation, as the contribution to $\hat{H}^{\Re}_{FU}$ vanishes (and, similarly, $E$ will also not enter into this process).
%
%
%
\section{Calculation of the single-spin dependent cross section}
\label{s:calccsFrag}
\begin{figure}[t] 
\begin{center}
\includegraphics[width=17cm]{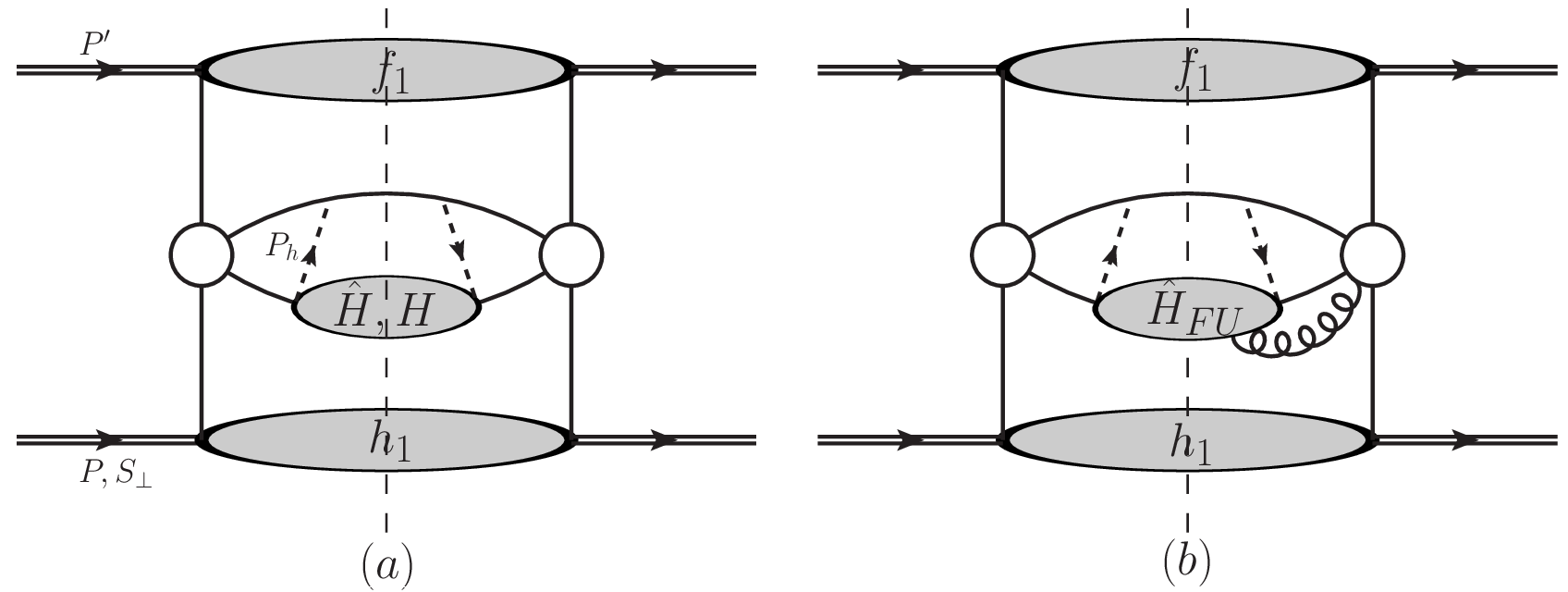}
\caption[] {Graphs showing factorization for the fragmentation contributions to $A_{UT}$ from (a) $qq$ correlators and (b) $qgq$ correlators.} \label{f:factFrag}
\end{center}
\end{figure}

The derivation of the fragmentation contribution to the SSA in $p^\uparrow p\rightarrow h X$ follows a similar procedure to what was outlined in Ref.~\cite{Metz:2012fq} for the double-spin asymmetry $A_{LT}$ in $p^\uparrow \vec{p}\rightarrow h X$.  The factorization of the reaction under consideration is shown in Fig.~\ref{f:factFrag}.  This includes collinear factors associated with the upolarized proton (top gray blob), the outgoing hadron (middle gray blob), and the transversely polarized proton (bottom gray blob) as well as hard factors (white blobs).  For each partonic channel the main task becomes calculating the hard scattering coefficients for each of these non-perturbative factors, which then allows us to write down the single-spin dependent cross section.  We will denote each channel by $ab\rightarrow cd$, where $a$$\,$($b$) is the parton associated with the transversely polarized (unpolarized) proton and $c$ is the parton that fragments into the detected hadron.

\begin{figure}[t] 
\begin{center}
\includegraphics[width=14cm]{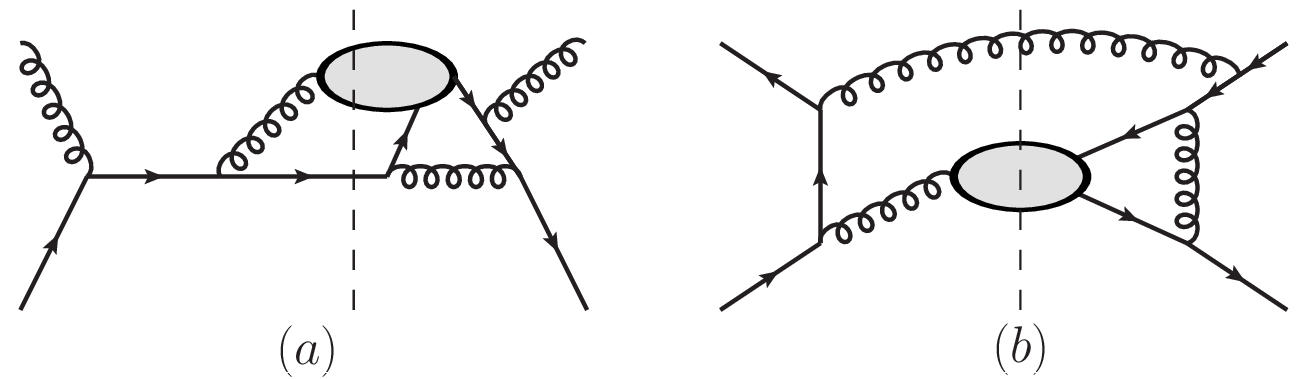}
\caption[] {Sample diagrams for (a) $qg$ and (b) $q\bar{q}$ induced channels where both quarks entering the fragmentation are on the same side of the cut.  Such graphs cancel after one sums all contributions.} \label{f:addgraphs}
\end{center}
\end{figure}

If we keep the transverse momentum (relative to the outgoing hadron) of the parton $c$ (as in Fig.~\ref{f:T3matrixFrag}(b)), then we can determine the hard scattering coefficient for $\hat{H}$ and its derivative.  On the other hand, if we neglect the transverse momentum of $c$ in the initial state (as in Fig.~\ref{f:T3matrixFrag}(a)), then we obtain the hard part for $H$.  Lastly, we must attach gluons to the hard blobs, in which case we can neglect the transverse momentum of the partons (as in Fig.~\ref{f:T3matrixFrag}(d)).  These diagrams allow us to find the $qgq$ contributions to the hard factors for $\hat{H}^{\Im}_{FU}$.  Note that potential contributions from graphs where the quarks entering the fragmentation are on the same side of the cut, like those in Fig.~\ref{f:addgraphs}, cancel after one sums the various diagrams, as we have shown through explicit calculation.  Such topologies can arise only through $qg$ and $q\bar{q}$ induced channels.  The $qgq$ diagrams can have internal parton lines go on-shell, in which case we employ the identity
\begin{equation}
\frac{1}{x \pm i\epsilon}=PV \frac{1}{x} \mp i \pi \delta(x)\,. 
\end{equation} 
As was the case in \cite{Liang:2012rb, Metz:2012fq}, only the $PV$ part survives the sum over the various cut diagrams.  These $qgq$ hard parts contain both $z_{1}$-dependent and -independent terms.  The latter can be pulled out of the integral over $z_{1}$, which allows $\hat{H}^{\Im}_{FU}$ for those pieces to be rewritten in terms of $\hat{H}$ and $H$ (using Eqs.~(\ref{e:DHhatFIm}), (\ref{e:EOMH})).  In the end, we find the fragmentation contribution to the cross section relevant for the SSA $A_{UT}$ in $A^\uparrow B\rightarrow C X$ is given by
\begin{eqnarray}
\frac{P_{h}^{0}d\sigma(\vec{S}_{\perp})} {d^{3}\vec{P}_{h}} \!\!\!&=&\!\!\! -\frac{2\alpha_{s}^{2}M_{h}} {S} \epsilon_{\perp,\alpha\beta}\,S_{\perp}^{\alpha}P_{h\perp}^{\beta}\sum_{i}\sum_{a,b,c}\int_{z_{min}}^{1}\frac{dz} {z^{3}} \int_{x'_{min}}^{1}\frac{d x'} {x'}\, \frac{1} {x}\,\frac{1} {x' S+T/z}\,\frac{1} {-x'\hat{t}-x\hat{u}} \,h_{1}^{a}(x)\,f_{1}^{b}(x')\nonumber\\ 
&&\hspace{-0.15cm}\times\,\left\{\left[\hat{H}^{c}(z)-z\frac{d\hat{H}^{c}(z)} {dz}\right]\,S_{\hat{H}}^{i} + \frac{1} {z} H^{c}(z)\, S_{H}^{i}\right.\nonumber \\[-0.1cm]
&& \hspace{5cm}\left.+\, 2z^2\int \frac{dz_1} {z_1^2} PV\frac{1} {\frac{1} {z}-\frac{1} {z_{1}}} \hat{H}_{FU}^{c,\Im}(z,z_{1})\,\frac{1} {\xi} \,S_{\hat{H}_{FU}}^{i}\right\},
\label{e:sigmaFrag}
\end{eqnarray}
where $i$ denotes the channel, $x=-x'(U/z)/(x'S+T/z)$, $x'_{min}=-(T/z)/(U/z+S)$, and $z_{min}=-(T+U)/S$.  Recall the Mandelstam variables $S$, $T$, $U$ and their respective partonic versions $\hat{s}$, $\hat{t}$, $\hat{u}$ were defined after Eq.~(\ref{e:process}).  We have introduced $\xi = z/z_g$, where $1/z_{g}=1/z-1/z_1$, and understand $1/\xi$ to mean $PV(1/\xi)$.  The hard scattering coefficients $S^{i}$ are given in Appendix A.

A few comments are in order on the final result.  First, note that the number of possible channels reduces significantly due to vanishing Dirac traces that arise in certain partonic interactions.  Second, we remark that the ``compact'' form involving $\hat{H}$ and its derivative manifest in Eq.~(\ref{e:sigmaFrag}) has been achieved by writing the cross section in terms of $\hat{H}$, $H$, and $\hat{H}^{\Im}_{FU}$.  Such a structure has arisen in other spin-dependent cross sections \cite{Qiu:1991pp, Kouvaris:2006zy, Koike:2007rq, Metz:2012fq, Metz:2010xs, Kang:2011jw, Liang:2012rb}.  In particular, we point out that unlike Refs.~\cite{Metz:2012fq, Liang:2012rb}, the compact form in (\ref{e:sigmaFrag}) was obtained by writing the result using F-type functions rather than D-type functions.  We repeat that the derivative term was first computed in Ref.~\cite{Kang:2010zzb}.    

%
%
%
\section{Summary and outlook}
\label{s:sum}
In conclusion, we have calculated the fragmentation contribution to the SSA in hadron production from proton-proton collisions using collinear twist-3 factorization, including terms involving quark-quark and quark-gluon-quark correlators.  The derivative term in this context had been computed before \cite{Kang:2010zzb}.  We generalized the work in Ref.~\cite{Zhou:2009jm} to the fragmentation case and specified the relevant collinear twist-3 functions --- see also \cite{Yuan:2009dw, Kang:2010zzb}.  The structure of the calculation essentially follows the procedure outlined in \cite{Metz:2012fq} for the double-spin asymmetry $A_{LT}$ in $p^\uparrow \vec{p}\rightarrow hX$.  Moreover, we find a ``compact'' form for $\hat{H}$ and its derivative that has also arisen in other spin-dependent cross sections \cite{Qiu:1991pp, Kouvaris:2006zy, Koike:2007rq, Metz:2012fq, Metz:2010xs, Kang:2011jw, Liang:2012rb}.  

The fragmentation mechanism is a key piece to the transverse SSA in $p^\uparrow p\rightarrow hX$, and an estimate of its impact on the asymmetry had been considered previously \cite{Kang:2010zzb, Kang:2011ni, Anselmino:2012rq}.  However, in order to completely determine the size of this term using the collinear twist-3 formalism, a full study including quark-gluon-quark correlators will be necessary.  The current work is an important step in this investigation.  In the end, a global analysis involving several processes will be required in order to extract the twist-3 functions (both distribution and fragmentation) that enter into the SSA in $p^\uparrow p\rightarrow h X$.  In this~context the AnDY Collaboration has recently measured the SSA in $p^\uparrow p\rightarrow jet\, X$ \cite{Bland:2013pkt}, which gives direct access to collinear twist-3 PDFs without ``contamination'' from the fragmentation side.  Along these lines, a measurement of Drell-Yan or the detection of direct photons in $p^\uparrow p$ collisions through experiments at RHIC would also provide valuable information.  Even a large $P_{h\perp}$ measurement of the Sivers and Collins asymmetries in SIDIS at JLab12, COMPASS, or a future Electron-Ion Collider can give insight into the collinear twist-3 distribution and fragmentation mechanisms, respectively, (without competition of one with the other) that contribute to $p^\uparrow p\rightarrow h\, X$.  Given the recent ``sign mismatch'' crisis involving the ETQS function and the Sivers function \cite{Kang:2011hk}, such an intensive study is worthwhile in order to thoroughly understand the mechanism behind this observable and whether the collinear twist-3 framework can appropriately describe it. 
\\[0.5cm]
%
%
\noindent
{\bf Acknowledgments:}
We would like to thank Z.-B.~Kang and J.~Zhou for useful discussions with regards to Ref.~\cite{Kang:2010zzb}.  This work has been supported by the NSF under Grant No.~PHY-1205942.

\appendix
%
%
%
%
\section*{Appendix A: Hard scattering coefficients} \label{a:FragS}
Here we give the hard scattering coefficients $S^{i}$ found in Eq.~(\ref{e:sigmaFrag}).  We mention that the $SU(3)$ color factors depend on $N_{c}=3$.  
\vspace{0.5cm}

\underline{$qg\rightarrow qg$ channel}
\begin{align}
S_{\hat{H}} &= (x-x')\left[ \frac{\hat{s}^2+\hat{u}^2} {\hat{t}^2} - \frac{1} {N_{c}^2}\right]\\[0.3cm]
S_{H} & = -\frac{1} {2} \left\{\left[\frac{\hat{t}^3(x'\hat{t}-x\hat{u})-6x'\hat{t}^2\hat{u}(\hat{s}-\hat{u})+4\hat{u}^3(2x'\hat{t}-x\hat{s})} {\hat{t}^3\hat{u}}\right] + \frac{1} {N_{c}^2}\left[\frac{x'\hat{s}+(x-x')\hat{u}} {\hat{u}}\right]\right\} \\[0.3cm]
S_{\hat{H}_{FU}} &= \frac{1} {N_{c}^2-1}\left[\frac{\hat{s}(\hat{s}-\hat{u})(-x'\hat{t}-x\hat{u})} {\hat{t}^3}\right] + \frac{1} {N_{c}^2} \left[\frac{x\hat{s}} {\hat{t}}\right] +\frac{x'\hat{s}^3+(x-x')\hat{u}^3} {\hat{t}^2\hat{u}}
\end{align}
\\[0.5cm]
\underline{$qq\rightarrow qq$ channel}
\begin{align}
S_{\hat{H}} &= \frac{N_{c}^2-1} {N_{c}^2}(x-x')\left\{-\frac{\hat{s}\hat{u}} {\hat{t}^2} + \frac{1} {N_{c}}\left[\frac{\hat{s}} {\hat{t}}\right]\right\}\\[0.3cm]
S_{H} & = \frac{N_{c}^2-1} {N_{c}^2} \left\{\frac{\hat{s}(x'\hat{s}^2+(x-x')\hat{u}^2)} {\hat{t}^3} - \frac{1} {N_{c}}\left[\frac{\hat{s}(x'\hat{s}^2-(x-x')\hat{u}(\hat{t}-\hat{u}))} {2\hat{t}^2\hat{u}}\right]\right\} \\[0.3cm]
S_{\hat{H}_{FU}} &= -\frac{x'\hat{s}^2} {\hat{t}^2} + \frac{1} {N_{c}}\left[\frac{(x-x')\hat{s}} {\hat{t}}\right] - \frac{1} {N_c^2}\left[\frac{\hat{s}(2x'\hat{t}^2+x\hat{u}(\hat{t}-\hat{u}))} {\hat{t}^3}\right] - \frac{1} {N_c^3} \left[\frac{\hat{s}(x\hat{u}^2-x'\hat{t}^2)} {\hat{t}^2\hat{u}}\right]
\end{align}
\\[0.5cm]
\underline{$q\bar{q}\rightarrow q\bar{q}$ channel}
\begin{align}
S_{\hat{H}} &= \frac{N_c^2-1} {N_c^2} (x-x')\left\{-\frac{\hat{s}\hat{u}} {\hat{t}^2} + \frac{1} {N_c}\left[\frac{\hat{u}} {\hat{t}}\right]\right\} \\[0.3cm]
S_{H} & =  \frac{N_c^2-1} {N_c^2} \left\{\frac{\hat{s}(x'\hat{s}^2+(x-x')\hat{u}^2)} {\hat{t}^3} - \frac{1} {N_c} \left[\frac{x'\hat{s}(\hat{s}-\hat{t})+(x-x')\hat{u}^2} {2\hat{t}^2}\right]\right\} \\[0.3cm]
S_{\hat{H}_{FU}} &= -\frac{(x-x')\hat{s}\hat{u}} {\hat{t}^2} + \frac{1} {N_c} \left[\frac{x'\hat{s}} {\hat{t}}\right] + \frac{1} {N_c^2}\left[ \frac{\hat{s}(2x'\hat{t}^2-x\hat{u}(\hat{s}-2\hat{t}))} {\hat{t}^3}\right] - \frac{1} {N_c^3} \left[\frac{x\hat{s}^2 -(x-x')\hat{t}^2} {\hat{t}^2}\right]
\end{align}
\\[0.5cm]
\underline{$\bar{q}q\rightarrow q\bar{q}$ channel}
\begin{align}
S_{\hat{H}} &= -\frac{N_c^2-1} {N_c^3}(x-x')\\[0.3cm]
S_{H} & =  -\frac{N_c^2-1} {N_c^3}\left[\frac{x'\hat{s}+(x-x')\hat{u}} {2\hat{u}}\right] \\[0.3cm]
S_{\hat{H}_{FU}} &= \frac{1} {N_c}\left[\frac{x\hat{s}} {\hat{t}}\right] - \frac{1} {N_c^3} \left[\frac{x'\hat{s}^2 +(x-x')\hat{u}^2} {\hat{t}\hat{u}}\right]
\end{align}
\\[0.5cm]
\underline{$qq' \rightarrow qq' $ channel}
\begin{align}
S_{\hat{H}} &= \frac{N_c^2-1} {N_c^2}(x-x')\left[-\frac{\hat{s}\hat{u}} {\hat{t}^2}\right]\\[0.3cm]
S_{H} & = \frac{N_c^2-1} {N_c^2} \left[\frac{\hat{s}(x'\hat{s}^2+(x-x')\hat{u}^2)} {\hat{t}^3}\right]  \\[0.3cm]
S_{\hat{H}_{FU}} &= -\frac{x'\hat{s}^2} {\hat{t}^2} -\frac{1} {N_c^2}\left[\frac{\hat{s}(2x'\hat{t}^2+x\hat{u}(\hat{t}-\hat{u}))} {\hat{t}^3}\right]
\end{align}

\underline{$q\bar{q}' \rightarrow q\bar{q}' $ channel}
\begin{align}
S_{\hat{H}} &= S_{\hat{H}}^{\,qq'\rightarrow qq'}\\[0.3cm]
S_{H} & = S_{H}^{\,qq' \rightarrow qq'} \\[0.3cm]
S_{\hat{H}_{FU}} &= -\frac{(x-x')\hat{s}\hat{u}} {\hat{t}^2} + \frac{1} {N_c^2} \left[\frac{\hat{s}(2x'\hat{t}^2-x\hat{u}(\hat{s}-2\hat{t}))} {\hat{t}^3}\right]
\end{align}
\\[0.5cm]
Note that we can obtain the channels involving antiquark fragmentation by charge conjugating the above partonic processes $ab\rightarrow cd$.  The respective hard parts are given by
\vspace{0.5cm}

\underline{$\bar{a}\bar{b} \rightarrow \bar{c}\bar{d}$ channels}
\begin{align}
S_{\hat{H}} &= S_{\hat{H}}^{\,ab\rightarrow cd}\\[0.3cm]
S_{H} & = S_{H}^{\,ab \rightarrow cd} \\[0.3cm]
S_{\hat{H}_{FU}} &= -S_{\hat{H}_{FU}}^{\,ab \rightarrow cd}
\end{align}

\end{document}